\def\be {\begin{equation}}
\def\ee {\end{equation}}

\def\R{{\sf R}}

\documentstyle[preprint,eqsecnum,aps]{revtex}
\begin{document}
\tighten
\draft
\preprint{UCSBTH-95-25, gr-qc/9510023}
\title{Mass Superselection, Canonical Gauge Transformations, and 
Asymptotically Flat Variational Principles}
\author{Donald Marolf}
\address{Physics Department, The University of California,
Santa Barbara, California 93106} \date{October, 1995}
\maketitle

\begin{abstract}
The phase space reduction of Schwarzschild black holes by 
Thiemann and Kastrup  and by Kucha\v{r}
is reexamined from a different perspective on
gauge freedom.  This perspective introduces additional
gauge transformations which correspond to asymptotically nontrivial
diffeomorphisms. 
Various subtleties concerning 
variational principles for 
asymptotically flat systems are addressed which allow us to 
avoid the usual conclusion that treating such 
transformations as gauge implies the vanishing of 
corresponding total charges.  Instead, superselection rules are found
for the (nonvanishing)
ADM mass at the asymptotic boundaries.   
The addition of
phenomenological clocks at each asymptotic boundary is 
also studied and compared with the `parametrization clocks' of 
Kucha\v{r}.
\end{abstract}

\section{Introduction}

In the study of constrained canonical systems, the term `gauge
transformation' is typically taken to mean `transformation generated
by a constraint' (see \cite{1AB}).  
However, when considering the equations of 
motion for a system, gauge transformations are usually identified
through
the appearance of arbitrary functions of time in the set of solutions.
These notions of gauge may differ in several ways.  In particular, when
acting on the canonical data corresponding to a Cauchy slice
$\Sigma$ of a
spacetime $M$, the orbits of the canonical gauge transformations may
be smaller than the orbits of the covariant gauge transformations.
A much cited example (see, for example \cite{1AB,TR,TK1,TK2,KK,GKZ})
is the case of general relativity
on asymptotically flat spacetimes, in which the covariant gauge
group contains all diffeomorphisms of the manifold while the
constraint-generated gauge group contains only those transformations
that preserve the asymptotic reference frame.

In this work we use canonical methods to study asymptotically
flat gravity but
do not assume that the constraints define the gauge
transformations.  Instead, we treat the gauge transformations
as an additional physical input which must be correctly incorporated
into the canonical description of our system.  
The reasons for this are threefold, as we shall now describe.  

The first aspect is philosophical.  The mathematical
definition of a physical system always requires physical input and, 
for the cases below, we wish to include in our system physics
not captured by the usual treatment.  As a result, we choose a somewhat
different mathematical definition.

The second aspect is observational.  For the case of so-called
`large' gauge transformations (those which are not continuously
connected to the identity), it is in fact standard that their
inclusion (or not) into the formulation of the system is taken
as an additional physical input.    For our study of asymptotically
flat gravitating systems, the `additional gauge transformations'
are those that act nontrivially in the asymptotic regions and
we observe that, for
the appropriate choice of topology on the space of gauge
transformations, 
they are in fact disconnected from the identity.  
As a result, the treatment below may be
viewed as taking a set of large transformations to be gauge\footnote{For 
example, we may take
a topology in which a sequence of infinitesimal gauge transformations
converges to the identity only if their gauge parameters $\epsilon_i(x)$
are such that $\int d^nx \ r^m |\epsilon_i(x)^2|$ (for appropriate $n,m$)
vanishes in the limit of
large $i$.}.

The third aspect is that 
Dirac canonical methods  \cite{Dirac}
can be regarded as a special example of a
class of algebraic descriptions of gauge theories (see \cite{GP})
based on so-called `generalized Peierls algebras\footnote{All 
of these algebras induce the same
(standard) algebra on gauge invariants, but differ on gauge dependent
functions.}'
instead of the Poisson algebra.
Quantization schemes based on these methods have 
`generalized constraints' which act much like
the constraints of Dirac quantization.  Now, if our system is defined by
an action principle, the action also
defines a notion of gauge through its invariance under transformations
that are compactly supported in time.  In this case, the orbits
generated by the generalized constraints are always
contained in gauge orbits defined by the action principle.
However, the constraint generated suborbits
may be much smaller than the full orbits by an amount that
depends on the particular
scheme employed.  It follows that the generalized constraints do not 
always generate
all of the gauge.  We take this as an indication that constraints
and gauge transformations are in fact two related {\it but different}
inputs for the mathematical formulation of a physical system.

In the work below, we are interested 
in asymptotically flat general relativity and we 
take our `physical' notion
of gauge transformations to be the entire group of diffeomorphisms
of some manifold $M$.  In particular,  
 we consider spherically symmetric asymptotically flat gravity, but
include gauge transformations that do not preserve the asymptotic
structure.  Our task, then, is to do this consistently.

The reduced phase space of this system has been
thoroughly investigated by Thiemann and Kastrup 
\cite{TK1,TK2} and by Kucha\v{r} \cite{KK}
using only the constraint-generated gauge transformations.
By making use of the results of Kucha\v{r} \cite{KK}, it is 
straightforward to include the additional gauge transformations
which lead immediately to superselection rules for
the ADM masses in the asymptotic regions.  This is because, while the
ADM Hamiltonian does not vanish, it generates the new
gauge transformations and so commutes with every gauge
invariant.  With this perspective on gauge, the
ADM Hamiltonian is thus {\it not} helpful in addressing what is known
as `the problem of time in quantum gravity' \cite{Bryce,KK2}.

We perform this analysis in section \ref{wo} and show the
consistency of this approach with various subtle points involving 
variational principles for asymptotically flat gravitating systems.
In particular, we show how to avoid the conclusion of 
\cite{GKZ} that our notion of gauge requires the asymptotic 
charges to vanish.  The main point will be made in 
subsection \ref{tA}, which treats a model of the asymptotic
boundary without the interior spacetime.  The new physical input enters
only at this boundary and attaching the interior is largely a 
technical issue, which we postpone until subsection \ref{tB}.
In section \ref{with},
we study the case of 
spherically symmetric gravity coupled to asymptotic phenomenological
clocks.  This is done 
both to provide an additional example of our approach
and for comparison with the parametrization clocks of \cite{KK}. 
From these examples, the corresponding
results for full asymptotically flat gravity (and indeed any field
theory) should be clear.

Below, the reader will encounter a number of `Remarks' which are
separated from the main text.  These clarify various technical points, 
but may be skipped without loss of continuity.

\section{The isolated Schwarzschild Black Hole}
\label{wo}

In this section, we address the isolated primordial
Schwarzschild black
hole, extending  the analyses of Thiemann and Kastrup\cite{TK1,TK2} and 
Kucha\v{r}\cite{KK} to treat asymptotically nontrivial
diffeomorphisms as gauge.
We find the language and notation of
\cite{KK} convenient and use it below.

As stated in the introduction, we treat the gauge transformations
of our system as a physical input and choose it to be the entire
group of spacetime diffeomorphisms.  We will take
our system to be defined by a variational principle using the 
Einstein-Hilbert action exactly because this captures the desired
notion of gauge.
Recall then that the usual Einstein-Hilbert action
\begin{equation}
\label{EH}
S = \int_M d^nx \sqrt{-g} R + 2  \int_{\partial M} K  dS
\end{equation}
on a spacetime $M$ with boundary $\partial M$ is invariant under
all diffeomorphisms of $M$.  Such diffeomorphisms necessarily
map the boundary $\partial M$ onto itself, but need not act
trivially on any part of $\partial M$.  Here, $R$ is the scalar
curvature, $K$ is the trace of the extrinsic curvature on the
boundary, and $dS$ is the proper volume element on the boundary.

As we are interested in Schwarzschild
black holes, we consider the restriction of this action to
spherically symmetric geometries.  This also restricts the allowed
gauge transformations to those that preserve spherical symmetry and
amounts to a partial gauge fixing, but the details of this
process will not concern us here.

A simple form of this
restricted action was derived in \cite{KK}, and we will
take this as our starting point.   By passing to spherically
symmetric canonical variables ($g,\pi$), performing a 
canonical transformation to the new variables $M,P_M,\R,P_\R$, 
and using the Lagrange multipliers $N^M,N^\R$, Kucha\v{r}
arrives at the action
\begin{eqnarray}
\label{kkaction}
S &=& \lim_{\rho \rightarrow + \infty}  \Biggl[
\int_{T_1}^{T_2} dt \int_{-\rho}^{\rho} dr (P_M(r) \dot{M}(r) + P_\R
(r) \dot{\R}(r) - N^M(r) M'(r)
- N^\R(r) P_\R(r))  \cr &+& \int_{T_1}^{T_2}
dt (N^M(\rho) M(\rho) - N^M(-\rho)M(-\rho)),
\Biggr]
\end{eqnarray}
where a dot indicates a time derivative, a 
prime denotes $\partial/\partial r$, and
the quantity $M(r)$ can be interpreted on the constraint surface as
the black hole mass.
Note that, because we consider primordial
black holes, the `radial' coordinate $r$ takes values on the entire real
line and that our spacetime has two asymptotic regions.
Here, we have written the action in a slightly different form than 
in \cite{KK} in order to make it clear this action can be well
defined even when the 
asymptotic limits $\lim_{r \rightarrow \pm \infty}
N^M(r)$ of the lapse do not exist.

\subsection{A model of a boundary}

\label{tA}

The viewpoint expressed in the introduction affects the analysis
only at the asymptotic boundaries.  The remaining task, to connect
the boundaries to the interior spacetime, is largely a technical issue.
As such, the main idea can be clearly illustrated for a finite dimensional
system which models a component of the boundary. We analyze such a
system here, and then return the the technical complications produced
by the interior in \ref{tB} below.

We will in fact study our model system in two different ways, first
through the a variational principle and then through a canonical
description.  Following the philosophy stated in the introduction, 
both formulations of the system will be designed to be compatible with
our `physical' idea that all diffeomorphisms should be treated as gauge.
As a result, they will differ somewhat from the familiar treatments.

Our starting point will be the action
\begin{equation}
\label{finS}
S = \int_{T_1}^{T_2} dt ( P \dot{M} - NM).
\end{equation}
This $M$ roughly corresponds to the total mass of an asymptotically
flat system, $P$
is it's conjugate momentum (as in \cite{KK}), and $N$ plays the role
of the asymptotic value of the lapse.  Following this analogy, the
standard treatment \cite{TK1,TK2,KK} of asymptotically flat systems
specifies $N(t)$ as part of the boundary data since this is a part
of the complete 3-metric around the boundary of the 4-dimensional
spacetime.  

The action \ref{finS} is invariant under the transformation
\begin{eqnarray}
\label{finG}
\delta P(t) &=& - \epsilon (t) \cr
\delta M(t) &=& 0 \cr
\delta N(t) &=& \dot{\epsilon} (t) 
\end{eqnarray}
for any $\epsilon (t)$ such that $\epsilon (T_1) = \epsilon (T_2) =0$.
We wish to consider such `time reparametrizations' as gauge
transformations.
There is, however, a subtle point in considering
such general gauge transformations when our system is formulated through
a variational principle.  Recall that a variational principle consists
of an action functional together with a set of boundary conditions that
the variations of the field must respect.  These boundary
conditions must be chosen so that the stationary points of
the variational principle 
include all of the desired solutions.
If the gauge transformations do not preserve
the boundary conditions, they will not lead to multiple solutions 
of the variational principle and need not be interpreted
as gauge.  Thus, if $N(t)$ is fixed (as in the usual treatment), there
are no gauge transformations and $(M,P)$ forms a canonical pair
of invariants.  We will not follow this approach, but instead fix
only boundary data compatible with interpreting \ref{finG} as gauge.

On the other hand, unrestricted variation of $N(t)$ would impose $M=0$
as an equation of motion, a feature that we wish to avoid to discuss
interesting black holes.  Our solution is to impose only that
$\int_{T_1}^{T_2} N(t) dt$ be specified as boundary data.  
The transformations \ref{finG} preserve
$\int_{T_1}^{T_2} N(t) dt$ as is appropriate to their interpretation as gauge.
We will be interested only in the correspondingly gauge invariant
quantities\footnote{As pointed out by a referee, one can obtain similar
results
by adding a term ofthe form $k \left( \int_{T_1}^{T_2} N(t) dt - A
\right)$ to the action, where $k$ is a time independent Lagrange
multiplier and $A$ is a constant which is not to be varied.  However, 
we will not pursue this approach here.}. 

The effects of varying $N(t)$ subject to this constraint are most
easily seen by expanding $N(t)$ and $M(t)$ in Fourier series on the
interval $t\in (T_1,T_2)$.  The constant Fourier component of
$N(t)$ has been fixed, but the other components may be varied freely.
As a result, these variations vanish exactly when $M(t)$ is a
constant. Varying $N(t)$ subject to our constraint imposes only
$dM/dt = 0$, which is also the equation of motion obtained by
varying $P(t)$.  Thus, our boundary condition imposes no new
equations of motion.  This completes our study of the variational
principle for the model system.

We now turn to a canonical description of this system.
Note that $P,M$ are the only canonical degrees of freedom, and that
they are conjugate.  While the lapse $N$ is not a canonical variable, its
treatment can again be made clear by expanding $N(t)$ in Fourier
modes.  The constant component is fixed by the boundary conditions,
while the other components act as Lagrange multipliers in the usual
way.  Thus, $N$ is in no way a coordinate on the phase space, but its
values are related to those of the canonical coordinates through
the equations of motion.

Let us consider the transformation induced
by \ref{finG} on the phase space $\Gamma$ by
identifying $\Gamma$ with the canonical data at time $t$.  This is just
$\delta M = 0$, $\delta P = -\epsilon$, which is clearly generated by
taking Poisson bracket with $\epsilon M$.  As a result, all (canonical)
gauge invariants are functions of $M$.

\begin{itemize}

\item{\bf Remark 1:}  Although we have found a generator for our gauge
transformations, we see no a priori reason why such a generator need
always exist.  Indeed, in the setting of generalized Peierls
algebras\cite{GP} or Poisson algebras \cite{PH}, 
there are examples for which such a generator
does not exist (c.f., axial `gauge-breaking' for Maxwell fields).

\end{itemize}

It seems that the gauge invariant phase space is just one dimensional.  
As such, it
is clearly not a symplectic space; that is, $M$ has no conjugate
momentum.  This means that there are {\it no} invariants that generate
nontrivial canonical transformations.  In a sense then, each different
value of the mass parameter $M$ defines a separate phase
space\footnote{Technically, each value of $M$ corresponds to a
different symplectic leaf\cite{cqrep} of the reduced phase space.} and we
have a classical superselection rule\cite{cqrep} between different values
of the mass.   Similarly, the quantization of this reduced
phase space leads to a collection of superselected one dimensional
Hilbert spaces, one for each allowed value of the mass.

\subsection{The full Schwarzschild spacetime}

\label{tB}

We now return to our study of the full spherically symmetric
vacuum theory as described by \ref{kkaction}.
Note that this action is invariant under the transformations
\begin{eqnarray} \label{gauge}
\delta M(r,t) = 0  \ &&  \ \delta \R(r,t) = \epsilon_\R (r,t) \cr
\delta P_M(r,t) = \epsilon_M'(r,t) \  && \ \delta P_\R(r,t) = 0 \cr
\delta N^M(r,t) = \dot{\epsilon}_M (r,t) \ && \
\delta N^\R(r,t) = \dot{\epsilon}_\R (r,t)
\end{eqnarray}
for any (smooth) $\epsilon_M$ that vanishes at $t = T_1,T_2$ and any
(smooth) $\epsilon_\R$.  
The gauge transformations \ref{gauge}
correspond to diffeomorphisms of the spacetime volume enclosed 
between $t = T_1,T_2$ which map the boundaries of this region
onto themselves.  
We will be interested only in objects that are
invariant under all such transformations.

As in \ref{tA}, we will choose boundary conditions in a manner
appropriate to treating \ref{gauge} as gauge.
To do so, let us consider a general variation
$\delta N_M$ of $N_M$.  The corresponding variation
of the action is
\be
\label{Nvar}
\delta S = - \lim_{\rho \rightarrow \infty}  \Biggl[
\int_{T_1}^{T_2} \int_{-\rho}^{\rho} dr  \ 
\delta  N^M(r)  M'(r) - \int_{T_1}^{T_2} dt (\delta N^M(\rho) M(\rho) - 
\delta N^M(-\rho) M(-\rho))  \Biggr].
\ee
Requiring $S$ to be stationary under such a general $\delta N$ would
impose $M'(r) = 0$, 
$\lim_{r \rightarrow \pm \infty} M(r) =0$ as 
equations of motion; i.e., $M(r) =0$.  To avoid this
difficulty, the lapse $N_M$ is typically taken (as in
\cite{TR,TK1,TK2,KK}) to be
fixed at spatial infinity. Note that this $N_M$ is simply part of the
3-metric on the (timelike) boundaries whose specification as boundary
data makes the usual variational principle well-defined.  This leads to the 
usual equations of motion, but removes the gauge transformations
\ref{gauge} with $\lim_{r \rightarrow \pm \infty} \epsilon_M(r) 
\neq 0$ from the class of allowed variations.  As a result, 
they are not typically treated as gauge.

This is not, however,  the only possible solution.
Since the variation of the action 
vanishes exactly under $\ref{gauge}$ whether or
not $\lim_{r \rightarrow \pm \infty} \epsilon_M(r) = 0 $, there
is no reason to specify the entire lapse on the boundary.  Instead, it
is only necessary to specify the lapse `up to a diffeomorphism of the
boundary.'  In particular, we may choose to fix not
$N_\pm (t) = \mp \lim_{r \rightarrow \pm \infty} N^M(r)$, 
but only the total proper time $\int_{T_1}^{T_2} dt N_\pm
(t)$ that elapses along the boundaries of our spacetime.  Then the variation
$\delta N_\pm (t)$ would not be completely free; the freedom would only
be that contained in the arbitrariness of $\dot{\delta N}_\pm$, and these
additional variations enforce only $dM_\pm/dt = 0$, where $M_\pm
= \lim_{r \rightarrow \pm \infty} M(r)$.
Since this follows from the original equations of motion, the
variation of the action is still well defined (and vanishing)
on the usual space of solutions.

\begin{itemize}

\item{\bf Remark 2:}  
One may wish to ask that the action
in fact be {\it differentiable} with respect to the allowed class
of variations.  This is a stronger condition than that the vanishing 
of a variation be well defined and requires all boundary terms to
disappear in the variation \ref{Nvar}.  Since, however, the
variation of the action vanishes completely under \ref{gauge}, it 
is differentiable with respect to \ref{gauge} and
we may still add these to the class of variations allowed 
by Kucha\v{r} \cite{KK}.  
This results in a smaller class of variations than that allowed
above.  
Note in particular that \ref{gauge} requires the variations of $N_M$ to be
correlated with those of $P_M$ so that this remark does not contradict
the argument given below \ref{Nvar}.
Since the action is unchanged by these additional variations, 
this new variational principle leads to the same equations of motion 
as that of \cite{KK}.
For the purposes below, 
the use of this variational principle or of the 
one stated above lead to precisely the same conclusions.

\item{ \bf Remark 3:}
With either class of variations above, the boundary data need
only be given up to a diffeomorphism of the boundary.
Specifying only
an equivalence class of boundary data may be unfamiliar, but leads
to no problems.  For example, Ref.  \cite{part} uses boundary data defined
only up to an even larger equivalence relation.
\end{itemize}

We now turn to the (Dirac style) canonical description of this system.
Note that $P_M(r),P_R(r)$ are the canonical conjugates of $M(r)$
and $R(r)$ and that variation of the Lagrange multipliers $N_M$ and $N_R$
enforce the constraints
\be
\label{con}
M'(r) = 0, \ \ \ 
P_\R(r) = 0.
\ee
However, the boundary term in the action \ref{kkaction} 
leads to the much publicized fact (see 
\cite{Bryce,D2,ADM,MTW} and others) that
the Hamiltonian is not just a sum of constraints.  Instead, it is
given by
\be
H(N^M,N^\R)
 = \int dr (N^M(r) M'(r)  + N^\R (r) P_\R (r)) + N_+ M_+ + N_- M_-
\ee
where $N^M(r)$ and $N^\R (r)$ are now treated as specified functions
of $r$ and are chosen so that $N_\pm = \mp \lim_{r \rightarrow \pm
\infty} N^M(r)$ exist.  With the inclusion of 
the boundary term, the Hamiltonian
is a differentiable function on the phase space.  Taking
Poisson Brackets with the Hamiltonian generates equations of motion
which, together with the constraints  \ref{con}, completely specifies
the dynamics of our system.

Following the philosophy outlined in the introduction, we take
as gauge all of the transformations \ref{gauge}.  That is, we base
our notion of gauge directly on the action and do not take
it to be determined by the form of the constraints\footnote{It is,
however, 
guaranteed that the transformations generated 
by the constraints will always be gauge (see, for example,
\cite{HT,Henneaux}).}.   Following the usual canonical procedure, we are 
particularly interested in
the transformation under \ref{gauge}
of a function $A(M,P_M,\R,P_\R)$ on the
phase space.  Many of these transformations are generated by a phase
space function as follows.  

Note that when the integral
\begin{equation}
\label{gen}
G(\epsilon_M,\epsilon_\R) =  \int dr (- \epsilon_M{}'(r) M(r)  + 
\epsilon_\R (r) P_\R (r)) 
\end{equation}
converges, it is differentiable on the phase space and generates the
transformations
\be
\{G,A\} = - \int d^3x \bigl[ {{\delta A} \over {\delta P_M}} 
\epsilon'_M + {{\delta A} \over {\delta \R}}
\epsilon_\R \bigr].
\ee
If we think of the phase space as the space of
canonical data on some Cauchy surface $\Sigma$  in $M$, then this is
just the transformation of $A$ induced by the
gauge transformations \ref{gauge}.  Thus, when \ref{gen} converges,
$G(\epsilon_M, \epsilon_\R)$ generates gauge transformations.
When $\epsilon_M$ and $\epsilon_\R$ are such that \ref{gen} does
not converge, the corresponding transformation \ref{gauge} has no generator.  
This is, however, 
consistent with our viewpoint as expressed in Remark 1.

\begin{itemize}
\item{\bf Remark 4:} In studying asymptotically flat
gravity, a number of conditions are typically imposed on the rate
at which the canonical variables approach their asymptotic
values.  We shall not explicitly choose a set of fall-off
conditions here, but
our viewpoint on this is the
same as for the other boundary data:  in principle, the fall-off conditions
need only be imposed {\it up to diffeomorphism (gauge) 
equivalence}, in which
case our transformations preserve the allowed class of data.  If
the fall-off conditions are not imposed in a diffeomorphism invariant
manner, then this is tantamount to gauge fixing and the system must
be reinterpreted accordingly.

\item{\bf Remark 5:} If one insists that all gauge transformations be
generated by phase space functions, one may impose fall-off
conditions such as $M(r) \rightarrow const$ and $\int_0^r dr P(r)
\rightarrow const$ (where neither constant is specified in the boundary
data).  The integral \ref{gen} then converges whenever $\epsilon_M (r)
\rightarrow const$ and $\epsilon_\R (r) \rightarrow const$.  The
corresponding class of gauge transformations is large enough to imply
all of our results below.

\end{itemize}

Note that $G$ is not just a combination of the constraints.
In fact, $G(\epsilon_M,\epsilon_\R)$ is identical to $H(\epsilon_M,
\epsilon_\R)$, the Hamiltonian evaluated at $N^M = \epsilon_M$, 
$N^\R = \epsilon_\R$.   Thus, any gauge invariant
phase space function
must Poisson commute with $H(N^M,N^\R)$ and,
while it is true that the ADM Hamiltonian is nonvanishing, 
it generates only the trivial transformation on the algebra of
gauge invariants.  With our perspective toward gauge, we find that
the ADM Hamiltonian is of
little use in solving `the problem of time' \cite{Bryce,KK2}. 

What then are the physical degrees of freedom of this system?
We answer this question by finding the complete set of gauge
invariant functions on the constraint surface.  From the form of
$G(\epsilon_M,\epsilon_\R)$, it is clear that any such
invariant $A$ must be independent of the momenta $P_M(r)$ and the
coordinates $\R (r)$.  Furthermore, on the constraint surface, 
the momenta $P_\R (r)$ vanish and $M(r)$ is independent of $r$.
Thus, the only gauge invariants are functions of the $r$-independent
quantity $M$, which is just the total mass of the system.
This coincides with the answer that would be expected from
Birkhoff's \cite{MTW} theorem:  every spherically symmetric solution 
to the vacuum Einstein equations is a piece of a Schwarzschild
spacetime, characterized only by its mass $M$.  
As in \ref{tA}, we find a superselection rule for the only
invariant degree of freedom, the total mass.

\section{Asymptotic Phenomenological Clocks}
\label{with}

For comparison, we now wish to study the system in which 
two phenomenological clocks are added to the black hole, one in
each asymptotic region. By phenomenological clocks, we mean
subsystems whose mathematical description is concocted to
ensure that their configuration variables $\tau$ increment 
in direct proportion to the passage of proper time along their
worldlines and such that this constant of proportionality is
independent of the state of the clock.  Nevertheless, we take the
clock to have a canonical description of the usual type, with
both a coordinate and a conjugate momentum (so that its 
phase space is a symplectic space).   
We shall also take the clock to be fixed (at $r = \pm \infty$)
and to ignore
any degrees of freedom associated with its movement.
We expect that
such a description could be approximated by realistic systems
in certain regimes, though we shall not discuss such approximations
here.  

In a fixed background $M$ foliated by slices
$\Sigma_t$ orthogonal to the clock's worldline, 
the standard description of a phenomenological clock 
is through the action
principle $\int dt \  P_\tau \dot{\tau} - NP_{\tau}$ where $N$ is the
lapse function defined at the clock's position
by the foliation.  If we were to couple this clock to the finite
dimensional model of \ref{tA}, the resulting action would be
\begin{equation}
S = \int_{T_1}^{T_2} dt ( P \dot{M} + P_\tau \dot{\tau} - N(M + P_\tau).
\end{equation}
Again imposing $\int_{T_1}^{T_2} dt N = const$, the gauge
transformations are
\begin{eqnarray}
\delta P (t) = - \epsilon (t)&,& \delta \tau (t) = \epsilon (t) \cr
\delta M (t) = 0&,& \delta P_{\tau} (t) = 0 \cr
\delta N(t) = \dot{\epsilon} (t) &&
\end{eqnarray}
for $\epsilon (T_1) = \epsilon (T_2) = 0$.  There are now three
independent gauge invariants: $M$, $P_\tau$, and $P + \tau$ and the
invariant $M +  P_\tau$ is superselected.  Attaching an asymptotic
clock to the full black hole is somewhat more subtle, as we shall see
below.

To study the
full Schwarzschild system coupled to asymptotic clocks, we use the 
action 
\begin{eqnarray}
\label{clockaction}
S &=& \int_{T_1}^{T_2} dt \int_{-\infty}^{\infty} (P_M(r) \dot{M}(r) + P_\R
(r) \dot{R}(r) - N^M(r) M'(r)
- N^\R(r) P_\R(r)) \cr &+& \int_{T_1}^{T_2} dt (P_{\tau^-} \dot{\tau}^- +
P_{\tau^+}\dot{\tau}^+ + N_- P_{\tau^-} - N_+ P_{\tau^+}
- N_+ M_+ - N_-M_-).
\end{eqnarray}
where we have now assumed that the asymptotic values 
$N_\pm = \lim_{r \rightarrow \pm \infty} \mp N(r)$ exist and we
have implicitly fixed the foliation to be orthogonal to the 
worldlines of the (asymptotic) clocks.  
Thus, we have introduced
a partial gauge fixing, of which we shall have to take proper
account.

\begin{itemize}
\item{\bf Remark 6:}
While it would have been nice to keep the full gauge symmetry
manifest, this is complicated by the asymptotic nature of the clocks.
The angles between the clocks' world lines and a $t = constant$
hypersurface is a function of the clock velocity $\dot{X}^\mu$,
the spatial metric, and the Lagrange multipliers.  Thus, to explicitly
display this (gauge) degree of freedom would require  the introduction
of  a clock
at finite position or the use of a new formalism to describe the velocity of
an asymptotic clock.  We choose to follow the conceptually
simpler gauge-fixed
approach here and we shall see that no problems arise as long as we
remember the physical setup behind this description.
\end{itemize}

Note that the action \ref{clockaction}
is exactly the same as that used by Kucha\v{r}
to describe `parametrization clocks.'  
However, we will require this action to be stationary
under a different class of variations;
we would like the addition of the clocks to affect the
gravitational degrees of freedom only through their explicit
interactions (as governed by \ref{clockaction}) and through the
gauge fixing that we have already performed. 
Since we have assumed
that the limits $\lim_{r \rightarrow \pm \infty} N^M(r)$ exist and 
that each $t= constant$ surface is orthogonal to the 
(asymptotic) clock worldlines, we require that the limits
$\lim_{r \rightarrow \pm \infty} N^M(r)$ exist and we impose
$\lim_{r \rightarrow \pm \infty} {\partial \over {\partial r}} N^M(r) =
0.$  Otherwise, however, we 
allow the same class of  variations for the gravitational
field here as when the clocks were absent (as in the
action \ref{kkaction}).  The clock fields $\tau^\pm$
are to be varied keeping $\tau^\pm(T_1), \tau^\pm(T_2)$
fixed while the clock momenta $P_{\tau^\pm}$ may be varied freely.

As before, we proceed by identifying the gauge transformations
associated with this action.  They are given by
\begin{eqnarray} \label{gaugeii}
\delta M(r,t) = 0 \ && \ \delta \R(r,t) = \epsilon_\R (r,t) \cr
\delta P_M(r,t) = \epsilon_M'(r,t) \ && \ \delta P_\R(r,t) = 0 \cr
\delta N^M(r,t) = \dot{\epsilon}_M (r,t) \ && \ \delta N_\R(r,t) =
\dot{\epsilon}_\R (t,r) \ \cr
\delta P_{\tau^\pm} (t) = 0 \  \cr
\delta \tau^\pm (t) = \mp \lim_{r \rightarrow \pm \infty} 
\epsilon_M(r,t).\ && \ \ \ \
\end{eqnarray}
where $\epsilon_M(T_1) = \epsilon_M(T_2) = 0$ and, as opposed to the 
case of \ref{gauge}, we must assume that $\lim_{r \rightarrow \pm
\infty} {\partial \over {\partial r}}\epsilon_M (r) = 0$ in order
to take account of our gauge fixing.
Note that the gravitational degrees of freedom transform just as they
did in \ref{gauge}.  

\begin{itemize}
\item{\bf Remark 7:} The detailed relationship between this
set of gauge transformations and the action \ref{clockaction}
is somewhat subtle, again due to the asymptotic nature of the clocks.
The action \ref{clockaction} is in fact invariant under
{\it all} transformations of the form \ref{gaugeii} for which
$\epsilon_M(t) = 0$ at $t = T_1,T_2$ whether or not 
$\lim_{r \rightarrow \pm \infty} {\partial \over
{\partial r}} \epsilon_M(r)$ vanishes or even exists.
However, we know that such a transformation breaks our gauge fixing
condition that the slices be orthogonal to the clock 
worldlines; in fact, a transformation of the form \ref{gaugeii}
for which ${\partial \over
{\partial r}} \epsilon_M(r,t) \neq 0$ at the clocks position
{\it would} have  changed the factor in the action that describes
the motion of the clocks.  (An additional subtlety is that the change
of this factor is only second order in $\epsilon_M$.)
Since we have dropped this factor from the
action, we must keep track of this point `by hand' through the
boundary condition ${\partial \over {\partial r}} N^M(r) = 0$
and the corresponding restriction ${\partial 
\over {\partial r}} \epsilon_M(r) =0$ on the gauge transformations.
\end{itemize}

We now proceed to the canonical formulation.  The constraints are
identical to those of section II (eq. \ref{con}).  The Poisson algebra
is also identical, with the addition of the new canonical
coordinates $\tau^\pm$ and their conjugates $P_{\tau^\pm}$.
The Hamiltonian is given by
\be
H(N^M,N^\R)
 = \int dr (N^M(r) M'(r)  + N^\R (r) P_\R (r)) + N_+ (M_+ +
P_{\tau^+}) + N_- (M_- - P_{\tau^-})
\ee
for $N^M$ such that $\lim_{r \rightarrow \pm \infty}
N^M(r)$ exists and $\lim_{r \rightarrow \pm \infty} {\partial
\over {\partial r}} N^M(r) = 0$.
Note that while $M(r)$ is still $r$-independent on the constraint
surface, the corresponding mass $M$ is no longer equal to the
ADM energy.  This is because the clocks were added on the
`outside' of the system and so contribute separately to the ADM
energy.  Indeed, we can now identify two distinct ADM energies
$E^\pm = M \pm P_{\tau^\pm}$, 
one at each asymptotic region.  We see that
these energies need no longer agree. 

Again, the gauge transformations can often be generated by taking Poisson
brackets with a phase space function; in this case, the function is
\begin{equation}
\label{g2}
G(\epsilon_M,\epsilon_\R) = \int dr (- \epsilon_M{}'(r) M(r)  + 
\epsilon_\R (r) P_\R (r))  - (\lim_{r \rightarrow  +\infty}
\epsilon_M(r)) P_{\tau^+} -  (\lim_{r \rightarrow  -\infty}
\epsilon_M(r)) P_{\tau^-}
\end{equation}
for $\epsilon_M(r)$ such that $\lim_{r \rightarrow \pm \infty}
\epsilon_M(r)$ exists and $\lim_{r \rightarrow \pm \infty}
{\partial \over {\partial r}} \epsilon_M(r) = 0$ and such that \ref{g2}
converges.  Again, we 
find that all gauge invariants commute with 
the Hamiltonian.  As mentioned above, the Hamiltonian in fact defines
two independent functions $E^\pm$ on the constraint surface, so this
will lead to {\it two} superselection rules, one for each asymptotic
region.

This is easily seen by explicitly carrying out the reduction.
Again, we have that $P_\R = 0$ and that gauge invariant phase
space functions are independent of $\R$.
However, 
because of the additional terms in the gauge generator, the
analysis of
the transformations parametrized by $\epsilon_M$ differs from that
of section \ref{wo}.
A gauge invariant need not be completely independent of $P_M(r)$, so
long as it depends on this momentum only through the combination
$P = \int_{-\infty}^{+\infty} dr P_M(r) + \tau^+ - \tau^-$.
The other gauge invariants are just the mass $M$ (as before) and
the two ADM energies $E^\pm$.  Since the $E^\pm$ are superselected, 
all of their Poisson brackets vanish while $\{M,P\} = 1$.
The observable $P$ has just the interpretation given in
\cite{KK} as the difference between the two clock
readings on any spacelike slice of constant killing time
through the black hole, 
and forms the conjugate
of $M$ as in \cite{KK}.  The corresponding quantum
theory is given by a set of superselected Hilbert spaces (labeled
by the values of $E^\pm$), each of which is isomorphic to
$L^2(\R^+,dm)$ on which $M$ acts by multiplication by $m$ and $P$
acts as $-i d/dm$.  Here we have incorporated the classical
positivity condition $M > 0$.
Note that by coupling clocks
we have introduced four canonical degrees of freedom $(\tau^\pm,
P_{\tau^\pm})$ and found three additional observables and one
new superselection rule.  

\begin{itemize}
\item{\bf Remark 8:}  In fact, due to our choice of gauge
fixing and boundary conditions (in particular, the restriction
$\lim_{r \rightarrow \pm \infty} {\partial \over {\partial r}}
\epsilon_M(r) = 0$), there are two other gauge invariants, 
$\lim_{r \rightarrow \pm \infty} P_M(r)$ which describe
\cite{KK} the asymptotic rate of change of Killing time
along the hypersurfaces.  Since our slices are asymptotically
orthogonal to the clocks' worldlines, $\lim_{r \rightarrow
\pm \infty} P_M(\infty)$ is a function of the angle between
these worldlines and the timelike Killing field of the Schwarzschild
metric.  As such, these invariants describe the
the velocities of the clocks with respect to the black
hole.  Since they lie at the boundaries, the $P_M(\pm \infty)$
commute with $M(r)$
and thus with all gauge invariants;  they too are superselected.
This is as it should be since we did not include the dynamics of the
clock velocities in \ref{clockaction}.  Note also that the
fall-off conditions of \cite{KK} would set $P_M(\pm \infty)$ 
to zero.  Here, we
have two options consistent with our philosophy:
\begin{itemize}
\item{1)} Impose that the fall-off conditions of \cite{KK} hold on
some Cauchy slice where the clocks read finite values.  This would set
$P_M(\pm \infty) = 0$.
\item{2)} Impose that the fall-off conditions imposed by
\cite{KK} on the metric and momenta hold up to a diffeomorphism of the
spacetime, but require no relation between slices satisfying
these fall-off conditions and the clock variables.  In this case, we may
have nonzero $P_M(\pm \infty)$.  This changes the above counting 
of degrees of freedom by adding two new invariants
$P_M(\pm \infty)$ which result from the
two restrictions 
$\lim_{r \rightarrow \pm \infty} {\partial 
\over {\partial r}} \epsilon_M (r) = 0$
placed on the gauge transformations.
\end{itemize}
\item{\bf Remark 9:}  Note that $P$ can be defined whenever
$P_M(+\infty) = -  P_M(-\infty)$ and $P_M \rightarrow P_M(\pm \infty)$
sufficiently rapidly.
\end{itemize}

We now present a brief comparison of this system to the 
black hole with `parametrization clocks' of \cite{KK}.
This type of clock is essentially defined by its appearance 
through the coordinate $\tau^\pm$ and the momenta $P_{\tau^\pm}$
in the variational principle used in \cite{KK}.
This variational principle uses the same action 
(\ref{clockaction}) as our phenomenological clocks but allows a larger
class of variations.  For the parametrization clocks, 
the lapse $N^M$ is allowed to be varied {\it freely} at spatial infinity.  
This has the effect of setting the ADM energies $E^\pm$
to zero.  Thus, `parametrization clocks' can be seen to be 
phenomenological clocks at rest with respect to the black hole (since
$P_M(\pm \infty) = 0$) and
whose internal states are chosen to carry
a negative energy that exactly cancels the mass-energy of the black
hole.

The `addition' of parametrization clocks to 
the black hole system is somewhat different from the usual notion of
coupling an additional system.  
Their inclusion modifies
the variational principle for the black hole not only by adding
new parameters to vary, but also by changing the class of allowed
variations for the lapse, a part of the metric field.
As a result, the counting of `additional' degrees of freedom is
more complicated than usual.  Using the notion of gauge transformation
of section \ref{wo}, the counting is just as that for the
phenomenological clocks, except that we are now also free to vary
$N_\pm$, imposing two new equations of motion ($E^\pm =0$).
As a result, the parametrization clocks create only one new
canonical degree of freedom.  Using the more limited notion of
gauge generated by constraints (as in \cite{KK}), they create
no new degrees of freedom.

One additional remark is that the free variation of $N^M$
is allowed only by the presence of the clocks {\it at}
the boundary.  Thus, if the clocks were first included at
a finite position (say, a fixed value of $\R$) and then
moved outward toward the boundaries, we would subject
the lapse to the boundary condition $\delta \int dt N_\pm(t) = 0$ 
as in section \ref{wo}.  There would then be no reason
for the total ADM mass to vanish and we would arrive at
the variation principle for the `phenomenological clocks.'

\section{Discussion}

As stated in the introduction, we consider the `proper'
set of boundary conditions and the definition of gauge transformations
to be a matter of physical input.  Clearly, in the case of
a gravitational system, there are two possible interpretations for a 
system with boundaries (asymptotic or not).  The first corresponds
to the treatment presented here, in which the system is considered
in isolation and no structure outside of these boundaries is utilized.
Another interpretation naturally yields the treatment of 
\cite{1AB,TK1,TK2,KK,GKZ,Bal}.  In this case, some observer
is assumed to sit at or just outside the boundaries but is
not explicitly included in the action functional.  This observer merely
supplies a coordinate chart on the boundaries (perhaps, through his
`parametrization clocks') which we may use to fix the gauge
of our system at the boundaries; for example, to fix
the lapse $N_\pm$.  If one wishes this external
observer to construct his clocks to yield zero ADM energy, we
have seen that he may do this as well.

However, if the first interpretation is chosen, we may naturally
take the full set of asymptotic transformations to be gauge.
It is clear that the results found above carry over to much more general
cases.  Fixing the boundary data for an asymptotically flat
variational principle only up to
a diffeomorphism of the boundary allows the full set of diffeomorphisms
of the spacetime volume to act as gauge symmetries of the action.
While this will not force the ADM energies to vanish, 
it will introduce a superselection law for the ADM energy
at each boundary.  If the requirement of spherical symmetry
is dropped, similar considerations apply to all of the
generators of the asymptotic Poincar\'e group.  As a result, none
of these generators will be gauge invariant themselves, but the
symplectic leaves of the reduced phase space (or the superselected
sectors of the quantum theory) will be labeled by the 
representations of the Poincar\'e group; for example by the mass 
$P_\mu P^\mu$.
In much the same way, consideration 
of these additional gauge transformations 
renders the quantities constructed in \cite{Bal} gauge dependent and 
implies that they generate pure gauge.

The idea of a superselection law for the mass of a gravitating system is
now new.  In particular, it was discussed in \cite{GKZ} and
\cite{HLM}, although
the motivations and arguments in these works were rather different both
from each other and from those presented here.  

The existence of a superselection 
rule for the total mass is quite in accord with our understanding of
other gauge theories, in which charge superselection is well 
known (see for example \cite{GKZ,Haag}).
However, our derivation
of this superselection rule (which could be applied to Yang-Mills
theories as well) is somewhat different from the usual 
one \cite{GKZ,Haag} given
for Yang-Mills theories, where the local (or `quasilocal') nature
of the observables is stressed.  Such a derivation is not applicable
here because, for diffeomorphism invariant
systems, gauge invariants can {\it never}
be local.  Given any function $f$ on the (unconstrained)
phase space built
from the fields in a compact
region $R$ of a spacetime $M$, 
we may simply perform a diffeomorphism that
moves $R$ into $M\setminus R$.  Since the the fields may vanish outside
$R$, only the
field independent (constant) function on $R$ can be
invariant under such a transformation. 
It is precisely due to the non-local nature of observables in general
relativity that great care is needed in formulating the gauge
transformations of the system.

\acknowledgments
The author would like to thank Abhay Ashtekar, Jorma Louko, Karel Kucha\v{r},
Jim Hartle, and Gary Horowitz for a number of useful discussions 
on these matters.  Special thanks go to Jorma Louko and Claudio
Teitelboim for providing the author's education in variational
principles, and again to Jorma Louko for critical comments
on an earlier draft.  This work was supported by NSF grant PHY90-08502.

\end{document}